\documentclass[fleqn,usenatbib]{mnras}
\usepackage{makecell}
\usepackage{newtxtext,newtxmath}
\usepackage{graphicx, float, hyperref, tabularx, physics, ulem}	
\usepackage{amsmath}	
\usepackage{amssymb}	
\usepackage{threeparttable}
\usepackage[T1]{fontenc}

\DeclareRobustCommand{\VAN}[3]{#2}
\let\VANthebibliography\thebibliography
\def\thebibliography{\DeclareRobustCommand{\VAN}[3]{##3}\VANthebibliography}

\usepackage{graphicx}	
\usepackage{amsmath}
\usepackage{url}
\usepackage{subcaption}	

\title[Timing Analysis of MAXI J1803$-$298]{
Timing Analysis of a New Black Hole Candidate MAXI J1803$-$298 with Insight-HXMT and NICER}

\author[Haifan Zhu et al.]{
Haifan Zhu,$^{1,2}$
Xiao Chen,$^{1,2}$
Wei Wang$^{1,2}$\thanks{E-mail: wangwei2017@whu.edu.cn}
\\
$^{1}$Department of Astronomy, School of Physics and Technology, Wuhan University, Wuhan 430072, China\\
$^{2}$WHU-NAOC Joint Center for Astronomy, Wuhan University, Wuhan 430072, China\\
}

\date{Accepted XXX. Received YYY; in original form ZZZ}

\pubyear{2023}

\begin{document}
\label{firstpage}
\pagerange{\pageref{firstpage}--\pageref{lastpage}}
\maketitle

\begin{abstract}
We present the timing analysis results of MAXI J1803$-$298, a black hole candidate, during its 2021 outburst using data obtained from the \textit{Insight-}HXMT and \textit{NICER} telescopes. Our analysis reveals that the source undergoes a state transition from the low/hard state to the hard intermediate state, followed by the soft intermediate state, and ultimately reaching the high/soft state. We searched for the quasi-periodic oscillations (QPOs) and studied the characteristics of the outburst. At the beginning of the outburst, the source was in the hard state, many type-C QPOs were seen in the \textit{Insight-}HXMT data, and the frequency of these QPOs increased from $\sim 0.16$ to $2.6$ Hz. Our analysis of the type-C QPOs' rms–frequency relationship indicates a turning point in the frequency. We also analyzed the phase lag versus frequency and energy relationship and deduced that the source likely has a high inclination angle, consistent with previous research. The observed rms and phase lag features in type-C QPOs could be explained by the Lense-Thirring precession model, while the alternatives would be still viable. The lag spectrum of type-B QPO exhibits a "\textit{U-shaped}" pattern similar to many other sources, and the type-B QPOs' rms increase as the energy rises. This phenomenon can be explained by the dual-corona model.  
    
\end{abstract}




\section{Introduction}

A black hole X-ray binary (BHXB) is a binary star system comprising a black hole (BH) and a regular companion star. Depending on their X-ray activity, they can be classified as either transient or persistent sources. Most black hole X-ray binaries spend the majority of their time in the quiescence phase, with low X-ray luminosity, and their X-ray emission becomes detectable only during outbursts. During an outburst, the X-ray emission from the system can vary by several orders of magnitude, and it typically displays distinct spectral states \citep{remillard2006}. The different spectral states of a BHXB can be broadly categorized as the low hard state (LHS), intermediate state (IMS), soft-intermediate state (SIMS), and high soft state (HSS).


During the initial phases of an outburst, the system is typically in the low hard state (LHS), where the X-ray spectrum is dominated by a nonthermal emission from an optically thin corona \citep{basak2016spectral}. A power-law component with a photon index of approximately 1.4-2.1 and a cutoff at high energies dominates the X-ray spectrum. As the outburst progresses, the system transitions to the high soft state (HSS), where the X-ray spectrum is dominated by a thermal component from the accretion disk, and the power-law component is either weak or nonexistent \citep{sharma2018spectral}. The intermediate state (IMS) is a transitional phase between the LHS and the HSS, and it can be further divided into the hard-intermediate state (HIMS) and soft-intermediate state (SIMS). During the IMS, the X-ray spectrum exhibits characteristics of both the LHS and the HSS.
At the end of an outburst, the X-ray flux from the black hole X-ray binary typically fades quickly, commonly by several orders of magnitude, as the accretion rate decreases, and the system returns to the LHS \citep{belloni2016astrophysics}. 

Quasi-periodic oscillations (QPOs) are often detected during the outburst phase of 
black hole X-ray binaries. It can be classified into two broad categories: low-frequency QPOs (LFQPOs) 
and high-frequency QPOs (HFQPOs). 
Based on their spectral and timing characteristics, LFQPOs in BHBs are  divided into 
type-A, type-B, and type-C categories. 
The quality factor  $ Q=\nu/\Delta \nu$ ($\Delta \nu$ is the full width at half maximum; FWHM), 
fractional root-mean-square (rms) variability, noise component, and phase lag are the main 
classification criteria \citep{casella2005abc,motta2015geometrical,motta2016quasi,ingram2019}.

Type-C QPOs have a central frequency range of approximately 0.01-30 Hz and are 
typically observed in the hard states of black holes, including both LHS and HIMS. 
They are characterized by high rms amplitudes of up to 20\%. 
type-C QPOs also exhibit strong band-limited noise, 
which means that the variability in X-ray brightness is concentrated within a certain 
range of frequencies. In addition, they usually exhibit second and sub-harmonics, 
which means that there are oscillations at twice and half the frequency of the main QPO \citep{ingram2019}. 

There has been a great deal of research done on type-C QPO origins. 
Most models used to explain type-C QPOs are either based on the disc's geometry or its inherent properties.
One of the proposed models for type-C QPOs is the relativistic precession model \citep{stella1997lense,stella1999correlations,schnittman2006precessing}, 
which suggests that the oscillations are caused by the inner accretion disc precession as a result of the frame-dragging effect of a spinning black hole. A variant precession model has been suggested as a possible explanation for the type-C 
QPOs saw in some accretion systems. 
This model suggests that a finite inner region of the accretion flow processes 
instead of a particle or a small ring. 
This model has the advantage of lowering the precession frequency 
to the right range and matches observations exceptionally well. As the inner precession region contracts and the precession frequency rises during the early phases of an outburst cycle, this also accounts for the observed rise in QPO frequency \citep{ingram2009low,ingram2011physical}. Furthermore, \cite{miller2005evidence} detected variations of the Fe K$\alpha$ line corresponding to the QPO phase in GRS 1915+105, indicating that the QPOs likely originate within the innermost regions of the disk. \cite{ingram2015phase} utilized phase-resolved spectroscopy and obtained similar results. Based on the analysis of the LFQPOs observed in H1743$-$322, it was discovered that both the energy of the iron line and the reflection fraction exhibit modulations corresponding to the QPO phase \citep{ingram2016quasi,ingram2017tomographic}.  These findings are in line with the theory that the Lense-Thirring precession causes the QPO, suggesting a geometric origin for the phenomenon. However, this model still has unresolved issues. Recent findings by \cite{marcel2021can} indicate that the Lense-Thirring mechanism neglects realistic accretion flow properties. Instead, they suggest that the extreme accretion speed and disk-driven ejections may play significant roles in producing type-C QPOs. 

Type-B QPOs are observed in black-hole X-ray binaries during the so-called Soft-Intermediate State (SIMS), which is a transitional state in the evolution of black-hole transients. The SIMS is defined by the presence of a type-B QPO \citep{belloni2016astrophysics}. A narrow peak, rms amplitude of up to 5\% and a centroid frequency of 5-6 Hz are the characteristics of type-B QPOs. These oscillations are frequently accompanied by weak red noise, which is typically present at a few percent rms or less and increases in amplitude at low frequencies (0.1 Hz). Along with the type-B QPO, a weak second harmonic and occasionally a subharmonic peak may be seen.

Although the exact origin of type B QPOs is still not fully understood, there are 
several models that explain the observed phenomenon of energy-dependent rms amplitude and lags well. A Comptonization model that considers a feedback loop between the hard and soft components of a LMXB was proposed by \cite{karpouzas2020comptonizing}. It is possible for the accretion disc or the surface of the compact object to act as the soft-photon source in that feedback loop, which involves Comptonized hard photons emitted by the corona and reflected back onto it. Later, at lower energies, the reflected photons are released. Using this model, \cite{garcia2021two} was able to fit the lag spectrum well and the rms spectrum poorly. Finally, they achieve excellent fits to the rms amplitude and phase lag spectra using two Comptonisation regions. \cite{bellavita2022vkompth} substituted the accretion disc for the compact object's surface as the soft-photon source to explain the characteristics of QPOs in black hole systems and successfully used it to explain the low-frequency QPOs. Recently \cite{peirano2023dual} suggest that the radiative properties of the type-B QPOs observed in GX 339-4 can be explained by this model. 

Black-hole candidate MAXI J1803$-$298 was detected by by the Gas Slit 
\textit{Camera of the Monitor of All-sky X-ray Image} (\textit{MAXI} /GSC;\citealt{matsuoka2009maxi})  
nova alert system on May 1, 2021. 
\cite{sanchez2022} presented a multiwavelength follow-up of the discovery outburst of the BH candidate 
MAXI J1803$-$298. During the initial hard state of an outburst, 
certain features are observed that suggest the presence of an outflow. 
Additionally, near-infrared emission lines show more complex profiles that may also be 
associated with outflows. These observations support the idea that there is a 
strong outflow during this outburst stage.
Furthermore, \cite{sanchez2022} proposed that the compact object observed in 
MAXI J1803$-$298 is likely a black hole with a mass range of approximately 3-10 $M_{\odot}$. 

According to \cite{jana2022}, a periodicity of approximately 7 hours 
was detected in the light curve of the object based on data obtained from the \textit{AstroSat}. The authors also estimated the mass function of the object to be $f(M) = 2.1 - 7.2 M_{\odot}$ and reported evidence of an evolving Compton corona and accretion disc. Additionally, the SIMS spectra revealed evidence of a disc wind. \cite{chand2022} also analyzed the MAXI J1803$-$298 during its outburst and found it to be in the hard-intermediate state, detecting QPOs and soft lags. 
Based on the X-ray spectroscopy, the source is likely a stellar mass Kerr black hole X-ray binary with a mass about 8.5-16 $M_{\odot}$ and spin $\gtrsim 0.7$. 
\cite{shidatsu2022} used \textit{MAXI}/GSC and \textit{Swift}/BAT data to study the X-ray spectral evolution of the black hole candidate MAXI J1803$-$298, revealing state transitions and flux variation. \cite{feng2022} used reflection models based on the data of \textit{Nuclear Spectroscopic Telescope Array (\textit{NuSTAR})} and \textit{the Neutron star Interior Composition Explorer ({NICER})}  to give the spin $\sim$ 0.991 and orbital inclination $\sim$ 70 degrees of this source.  \cite{coughenour2023reflection}, utilizing \textit{NuSTAR} observation data, presents evidence of a highly inclined, high-density accretion disk, potential type-B and type-C QPOs, and outflowing disk winds during its 2021 May outburst. \cite{wood2023time} could constrain the flux density variability and proper motions of three components of MAXI J1803$-$298, and inferred the ejection date of a transient jet during the peak of the outburst, using their new model fitting approach directly applied to interferometric observations.  

In this paper, we present the timing analysis of the source using Hard X-ray Modulation Telescope (\textit{Insight-}HXMT) and \textit{NICER} observations. 
Section~\ref{obs} describes the observations and data reduction methods. 
In Section~\ref{result}, we present our results. Sections~\ref{discussion} and \ref{con} presents the discussion and conclusions.

\section{OBSERVATIONS AND DATA REDUCTION}
\label{obs}
\subsection{Observations}
From May 3 to July 28, 2021, MAXI J1803-298 was observed by the Hard X-ray Modulation Telescope (\textit{Insight}-HXMT) during a new outburst detected by MAXI/GSC. \textit{Insight}-HXMT, China's first X-ray astronomy satellite, was launched on June 15, 2017, and has three detectors that cover different energy ranges: High Energy (HE) detectors ranging from 20.0 to 250.0 keV \citep{liu2020High}, Medium Energy (ME) detectors ranging from 5.0 to 30.0 keV \citep{cao2020medium}, and Low Energy (LE) detectors ranging from 1.0 to 15.0 keV \citep{chen2020low}, with areas of 5100 $\rm cm^2$, 952 $\rm cm^2$, and 384 $\rm cm^2$, respectively.

The Neutron star Interior Composition Explorer (\textit{NICER}; \citealt{gendreau2016neutron}) is an X-ray telescope on the International Space Station (ISS) that used silicon-drift detectors with a sensitivity range of 0.2 to 12 keV. The effective area of \textit{NICER} is >2000 $\rm cm^2$ at 1.5 keV and 600 $\rm cm^2$ at 6 keV, respectively, for the detectors. The instrument's time-tagging resolution is 300 nanoseconds. \textit{NICER} observed MAXI J1803-298 during its outburst from May 2 to November 7, 2021.

\subsection{Data Reduction}

The Insight-HXMT Data Analysis Software (HXMTDAS)\footnote{\url{http://hxmten.ihep.ac.cn/software.jhtml}} V2.04 was used to extract and analyze the data.
We generated our good time intervals by applying the following criteria: a pointing offset angle of less than 0.04°, an elevation 
angle greater than 10°, a geomagnetic cutoff rigidity larger than 8°, and exclusion of data within 300 seconds of the passage through the South Atlantic Anomaly (SAA). The light curves were generated using $helcgen$, $melcgen$, and $lelcgen$ tools in HXMTDAS. 
The background was estimated using HEBKGMAP, MEBKGMAP, and LEBKGMAP, and further 
processed using $lcmath$ to remove the estimated background. The energy ranges selected for our analysis are 1-10 keV (LE), 10-20 keV (ME), and 20-100 keV (HE). 

The primary objective of \textit{NICER} is to investigate the characteristics of neutron stars, specifically their dense cores and strong gravitational fields. In this study, we have employed X-ray data acquired by \textit{NICER} to examine the properties of MAXI J1803$-$298. To perform the data reduction, we used the NICERDASv10 software distributed with HEASOFT v6.31.1. The current version of calibration files, CALDB xti20221001, was released on October 31, 2022, along with the \textit{NICER} Calibration documentation. The \textit{NICER} data reduction pipeline, which consists of the \textit{nicerl2} and \textit{nicerl3-lc} tasks, was utilized to calibrate and generate light curves.
The energy range selection for \textit{NICER} was 1-10 keV, appropriate for studying the X-ray 
properties of our source.  
The low background flux estimates obtained using the \textit{nibackgen3C50} task  allowed us to neglect any background subtraction in the light curve analysis.

\subsection{Timing analysis methods}
To calculate the power density spectra (PDS) for each observation, we used the \textit{powspec} tool with a time interval of 64 seconds and a corresponding time resolution of 1/128 seconds. The PDS obtained for each observation were then averaged, and the resulting PDS was re-binned using a geometric factor of 1.05 in frequency space. We normalized the PDS to units of $\rm rms^2/Hz$ and subtracted Poisson noise \citep{belloni1990atlas, miyamoto1991x}. We estimated the total fractional variability (rms of the PDS) in the range of 0.01–32 Hz of LE. 

We utilized multiple Lorentz functions to fit the profiles of both the broadband noise (BBN) and QPO in the PDS with the QPO components that we obtained. This enabled us to extract the fundamental parameters of the QPO. To determine the fractional rms of the QPO, we calculated
\begin{equation}
\rm rms_{QPO}=\sqrt{R}\times\frac{S+B}{S},
\end{equation}
where S represents the source count rate, R represents the normalization of the QPO's Lorentzian component, and B represents the background count rate \citep{bu2015correlations}.

To examine the energy-dependent properties of the QPO, we partitioned the $Insight$-HXMT data into various energy ranges: 1-3 keV, 3-5 keV, 5-10 keV, 10-13 keV, 13-16 keV, 16-20 keV, 20-40 keV, 40-660 keV, and 60-100 keV. Similarly, for the \textit{NICER} data, we utilized the following energy bins: 1.0-1.5 keV, 1.5-2.0 keV, 2.0-3.0 keV, 3.0-4.0 keV, 4.0-6.0 keV, 6.0-8.0 keV, and 8.0-10.0 keV. Upon sorting the data into these energy bands, we noticed that the QPO signal vanished in some ranges. Consequently, we excluded these energy ranges and solely examined the bands where the QPO signal was present.

We calculated the cross spectra of the light curves obtained from the low-energy and high-energy bands. The phase-lags between the different energy bands were computed by taking the 1-3 keV band as the reference for $Insight$-HXMT and 2.0-3.0 keV for \textit{NICER}. The utilization of $2-3$ keV as the reference energy for NICER data analysis is similar to other work on BHXBs based on NICER \citep{bellavita2022vkompth,peirano2023dual}, so that the results would be compared with each other. Positive lags, or hard photons lagging behind soft photons, were referred to as hard lags. To determine the necessary phase-lags and other parameters, we employed the $stingray$ package. We calculated the phase lags of the QPOs over a frequency range centered on the QPO centroid frequency and dispersed over its FWHM. We estimated the uncertainties of the parameters using Markov Chain Monte Carlo (MCMC) simulations with 200 walkers and a length of $10^5$ steps. The uncertainties reported in this paper are given at the 90\% confidence level.

\section{RESULT}
\label{result}
\subsection{Light curves and HID of MAXI J1803$-$298 }
\begin{figure}
    \includegraphics[width=\columnwidth]{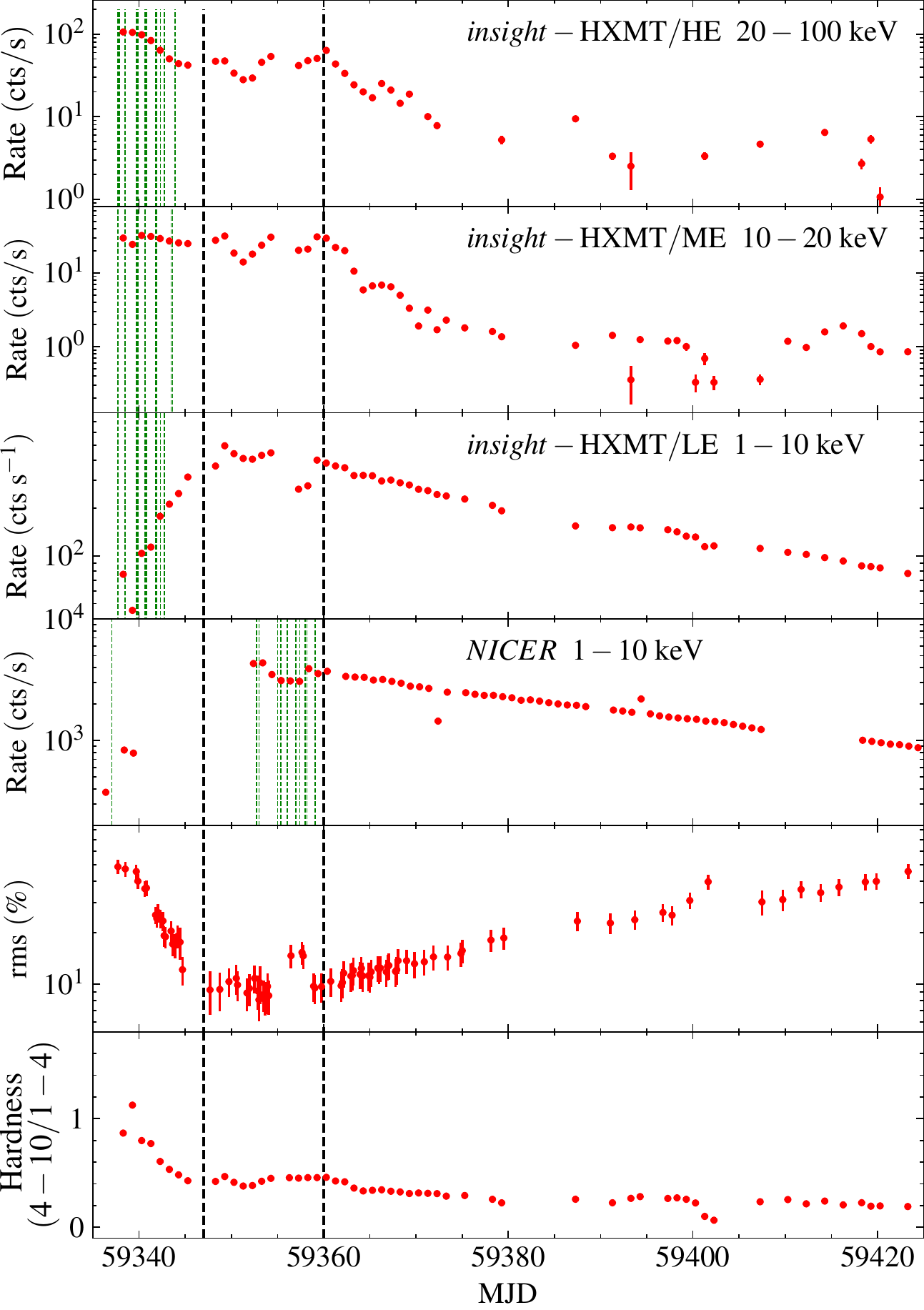}
    
    \caption{The light curve of MAXI J1803$-$298 is presented in the following order from top to bottom: 
    $Insight$-HXMT HE (20-100keV) data, $Insight$-HXMT ME (10-20keV) data, $Insight$-HXMT LE (1-10keV) 
    data, observational data from \textit{NICER} (1-10keV), the fractional rms of PDS in the range 0.001-32 Hz, 
    and  the hardness ratios calculated from the data obtained by $Insight$-HXMT with $\rm (4-10 keV)/(1-4 kev)$. 
    The vertical green dashed line in the figure denotes the QPO signal observed during this period, 
    the black dashed line represents the transition of the state.  }
    \label{figure1}
\end{figure}

After MAXI/GSC discovered the outburst of MAXI J1803$-$298, several X-ray telescopes observed it. 
Figure~\ref{figure1} shows the background-subtracted light curves from $Insight$-HXMT and \textit{NICER}, hardness ratios, and 
the rms of the PDS. 
The count rate of the HE band remained at its 
peak of approximately 106 $\rm cts~s^{-1}$ from the beginning, followed by a gradual decline. 
It maintained a relatively high count rate between MJD 59347 and MJD 59361 before rapidly dropping 
to nearly 0 afterward.	The count rate of the ME band exhibited similar behavior to that of the HE band. The 
count rate of the LE band, on the other hand, slowly increased from less than 100  $\rm cts ~s^{-1}$ at the beginning and reached its peak of 494 $\rm cts~ s^{-1}$ on MJD 59349. It remained stable for a long period before slowly declining after MJD 59361. 

The light curve data obtained by \textit{NICER} exhibited behavior similar to that of the LE data from $Insight$-HXMT. 
From the hardness ratios plot, it can be observed that the hardness rapidly decreased in the initial phase of the eruption. It remained stable for a considerable ti period after MJD 59347 and 
further softened after MJD 59361. 

At the beginning of the observation,  The fractional rms of PDS was at a relatively high level, during which a large number of type-C QPOs appeared. As time evolved, the rms value decreased to below 10\%, and remained stable for a long period, before rising back to a high level after MJD 59361. 

   \begin{figure}
    \includegraphics[width=\columnwidth]{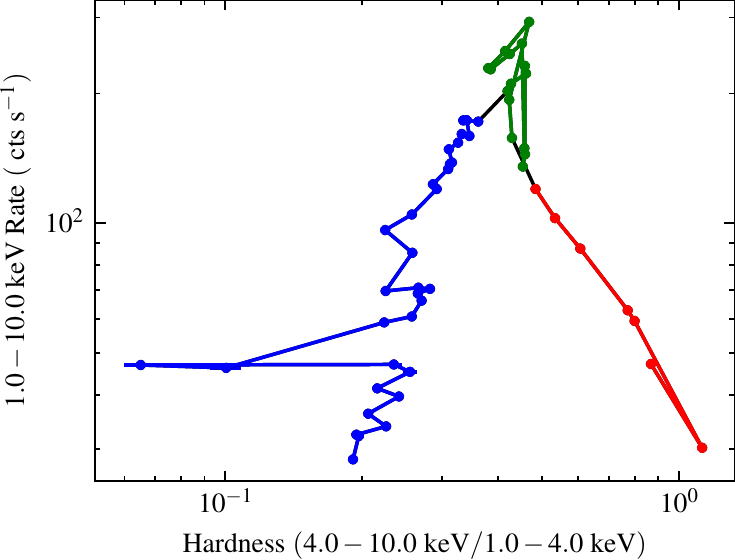}
    
    \caption{HID of MAXI J1803$-$298, the horizontal axis represents hardness ratios, 
    with data obtained from $Insight$-HXMT LE (4-10 keV/1-4 keV). The vertical axis represents the count rate of $Insight$-HXMT LE (1-10keV).
    The color red represents the low hard state, green represents the intermediate state, and blue represents the soft state. }
    \label{figure2}
   \end{figure}
 
In Figure~\ref{figure2}, we have plotted the HID of MAXI J1803$-$298, which exhibits a similar evolutionary path to other BHXBs such as MAXI J1535$-$571 \citep{zhang2022}. The outburst commences at the LHS, which corresponds to the lower right corner of  Figure~\ref{figure2}, where the fractional rms persists at an elevated level exceeding 40\%. As the counting rate increases, the rms remains at $\sim (20-30)\%$, and the corresponding hardness ratio in Figure~\ref{figure2} experiences a rapid drop during this phase. Subsequently, the system progresses towards the upper left of Figure~\ref{figure2}, and the rms value decreases to approximately 10\%, while the hardness ratio remains constant after MJD 59349, indicating that the system has entered the IMS. At MJD 59352, type-B QPOs were detected in the corresponding PDS of \textit{NICER}.  This suggests that the source was in the SIMS, but its precise classification could not be accurately determined as \textit{NICER} did not observe before this period. After MJD 59361, the source began to shift towards the lower left of the HID plot, and the fractional rms of PDS started to increase. This indicates that the source has entered the HSS. 

\begin{figure*} 
    \includegraphics[width=0.4\textwidth]{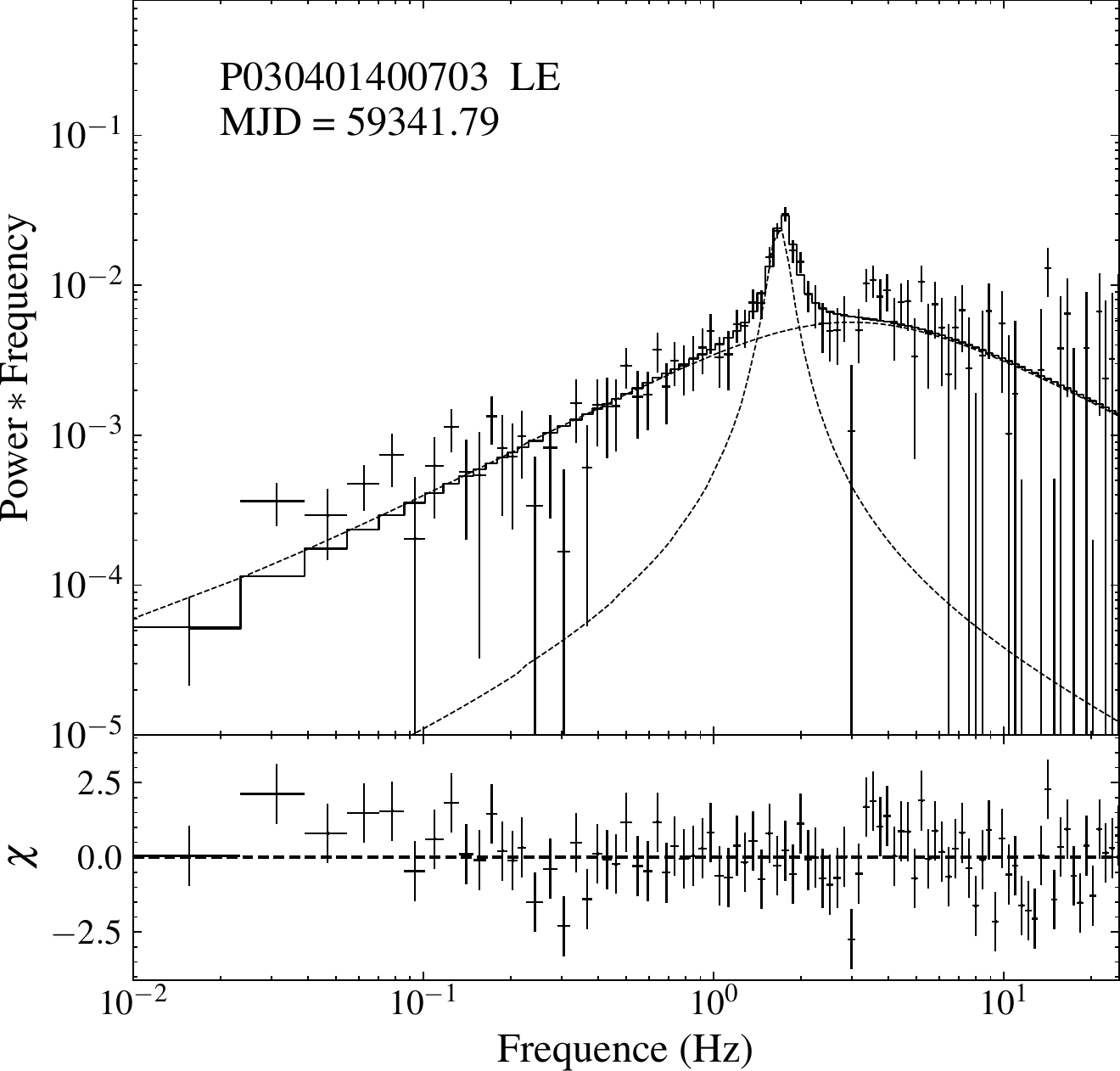}$~~~~$
    \includegraphics[width=0.4\textwidth]{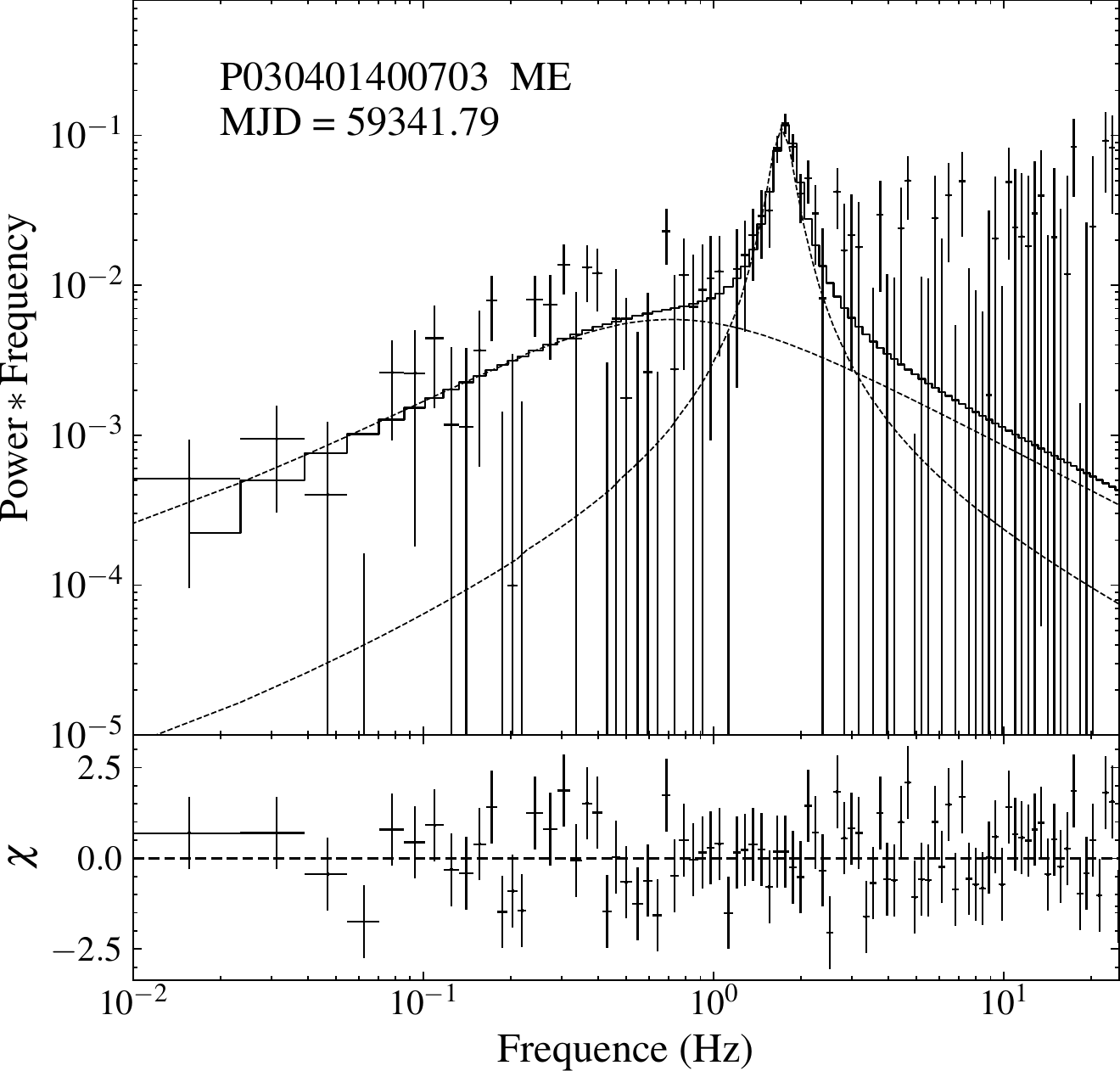}
    \includegraphics[width=0.4\textwidth]{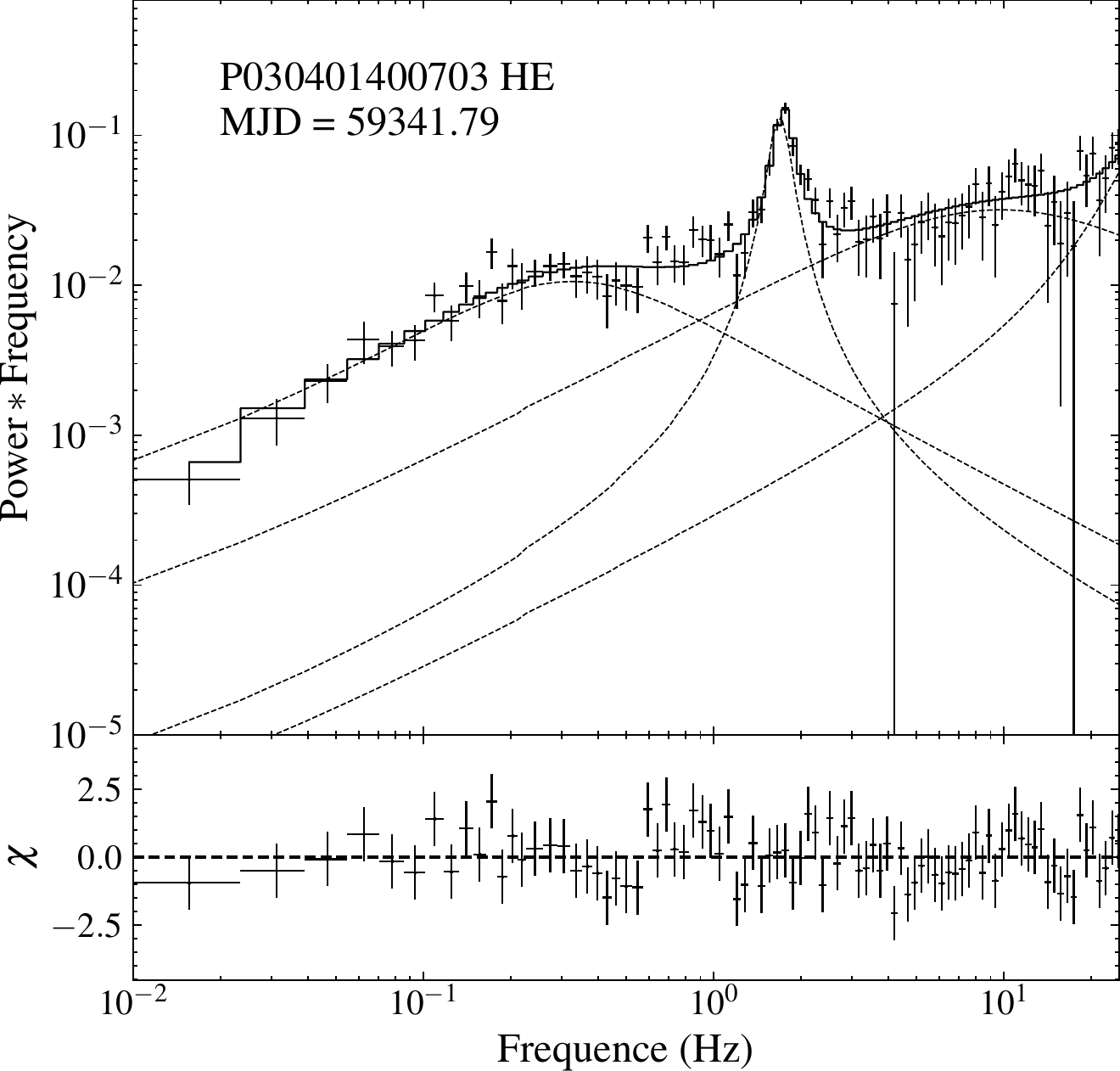}$~~~~$
    \includegraphics[width=0.4\textwidth]{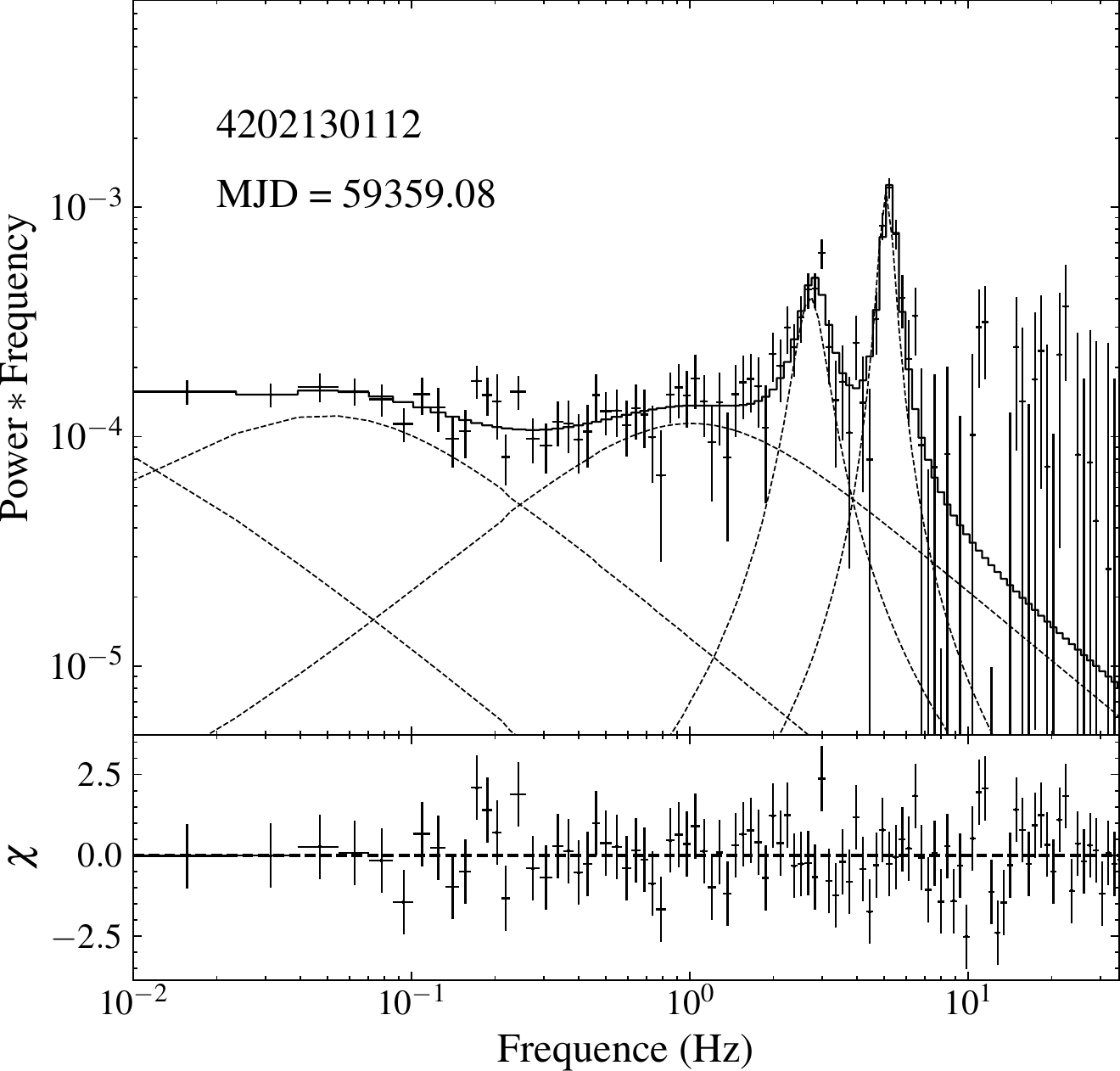}
    \caption{The representative fitting results of the QPO signals were obtained using data from 
    \textit{insight-}HXMT LE, ME, HE, and \textit{NICER} (right bottom panel). 
    The ObsID and MJD are labeled in the respective figures.} 
    \label{hxmtpds}
\end{figure*}

\subsection{Type-C QPOs} 
In Figure~\ref{hxmtpds}, we present the PDS obtained from the three detectors of the $Insight$-HXMT, as well as those obtained from \textit{NICER}, accompanied by their respective model fitting results. We present our analysis results for type-C QPOs first, and the results for type-B QPOs will be presented in the next subsection.

\begin{figure}
    \includegraphics[width=\columnwidth]{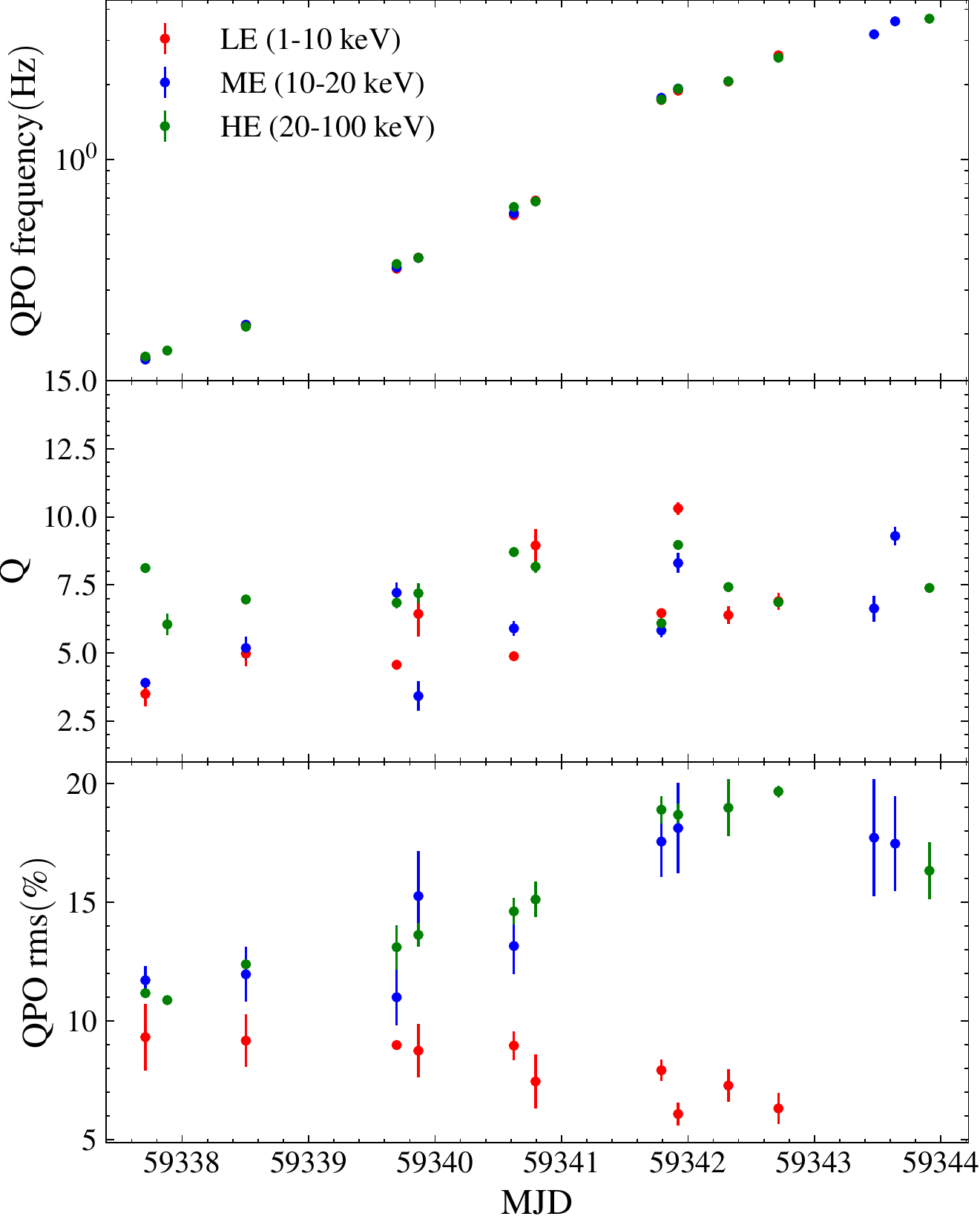}
    \caption{The evolution of the frequency, $Q$ value, and rms of the type-C QPOs with time. The data are obtained from three different $Insight$-HXMT detectors: HE (20-100keV, blue), ME (10-20keV, green), and LE (1-10keV, red). 
    All three figures have the same legend.
     }
    \label{figure3}
   \end{figure}

Table~\ref{tab:qpo} summarizes the fitting results of type-C QPO parameters for all data obtained by the LE of $Insight$-HXMT, including centroid frequency ($\nu$), $Q$, and the QPO rms. Figure~\ref{figure3} displays the evolution of type-C QPOs obtained from data obtained by $Insight$-HXMT. Moreover, type-C QPOs were also detected in the early phase of the outburst in the observational data obtained from \textit{NICER} (refer to Figure~\ref{figure1}).
However, due to the lack of continuous observations and the close proximity of the \textit{NICER} observation time to that of the $Insight$-HXMT, we mainly utilized the observation data from $Insight$-HXMT for the analysis of the type-C QPOs and only presented the $Insight$-HXMT results. 

\begin{table}
    \caption{Observations of type-C QPOs in MAXI~J1803$-$298. The given error represents a 90\% confidence level. The presented fitting results are derived from the LE data. }
    \centering
    
    \begin{threeparttable}
        \centering
    
    \begin{tabularx}{\columnwidth}{ccccc} 
        \hline
        ObsID\tnote{a} & MJD & $\rm QPO$\tnote{b}$\rm (Hz)$ & $\rm Q$&$\rm rms (\%)$ \\
        \hline
        101 & 59337.71 & $0.161\pm 0.008$ & $3.49\pm 0.45$&$9.32\pm 1.41$ \\
        201 & 59338.50 & $0.217\pm 0.004$ & $4.97\pm 0.46$&$9.17\pm 1.10$ \\
        301 & 59339.70 & $0.365\pm 0.003$ & $4.56\pm 0.06$&$8.98\pm 0.08$\\
        302 & 59339.87 & $0.404\pm 0.006$ & $6.43\pm 0.84$&$8.75\pm 1.11$ \\
        401 & 59340.62 & $0.598\pm 0.009$ & $4.88\pm 0.17$&$8.96\pm 0.61$ \\
        402 & 59340.79 & $0.684\pm 0.012$ & $8.95\pm 0.59$&$7.45\pm 1.13$ \\
        703 & 59341.79 & $1.731\pm 0.015$ & $6.46\pm 0.18$&$7.92\pm 0.45$ \\
        704 & 59341.92 & $1.887\pm 0.021$ & $10.31\pm 0.22$&$6.08\pm 0.48$ \\
        707 & 59342.32 & $2.052\pm 0.033$ & $6.39\pm 0.33$&$7.28\pm 0.70$ \\
        710 & 59342.71 & $2.614\pm 0.049$ & $6.90\pm 0.31$&$6.31\pm 0.65$ \\
        \hline
    \end{tabularx}
    \begin{tablenotes}
        \item[a] Obs IDs: P030401400NNN
        \item[b] QPO represents the centroid frequency of QPOs
    \end{tablenotes}
\end{threeparttable}
\label{tab:qpo}
\end{table}

Significant type-C QPO signals were detected in all three Insight-HXMT detectors during the early phase of the outburst. No apparent harmonic signal was observed, and the QPO frequency gradually increased over time, rising from approximately 0.16 Hz to around 2.6 Hz during the entire hard state. In terms of the QPO rms evolution, the trend observed in the LE detector differed from that of the other two detectors. The rms remained relatively stable during the low-frequency period. As the frequency increased over time, the rms of the LE detector gradually decreased to about 6\%, while the rms of the ME and HE detectors increased gradually.

\begin{figure}
    \includegraphics[width=\columnwidth]{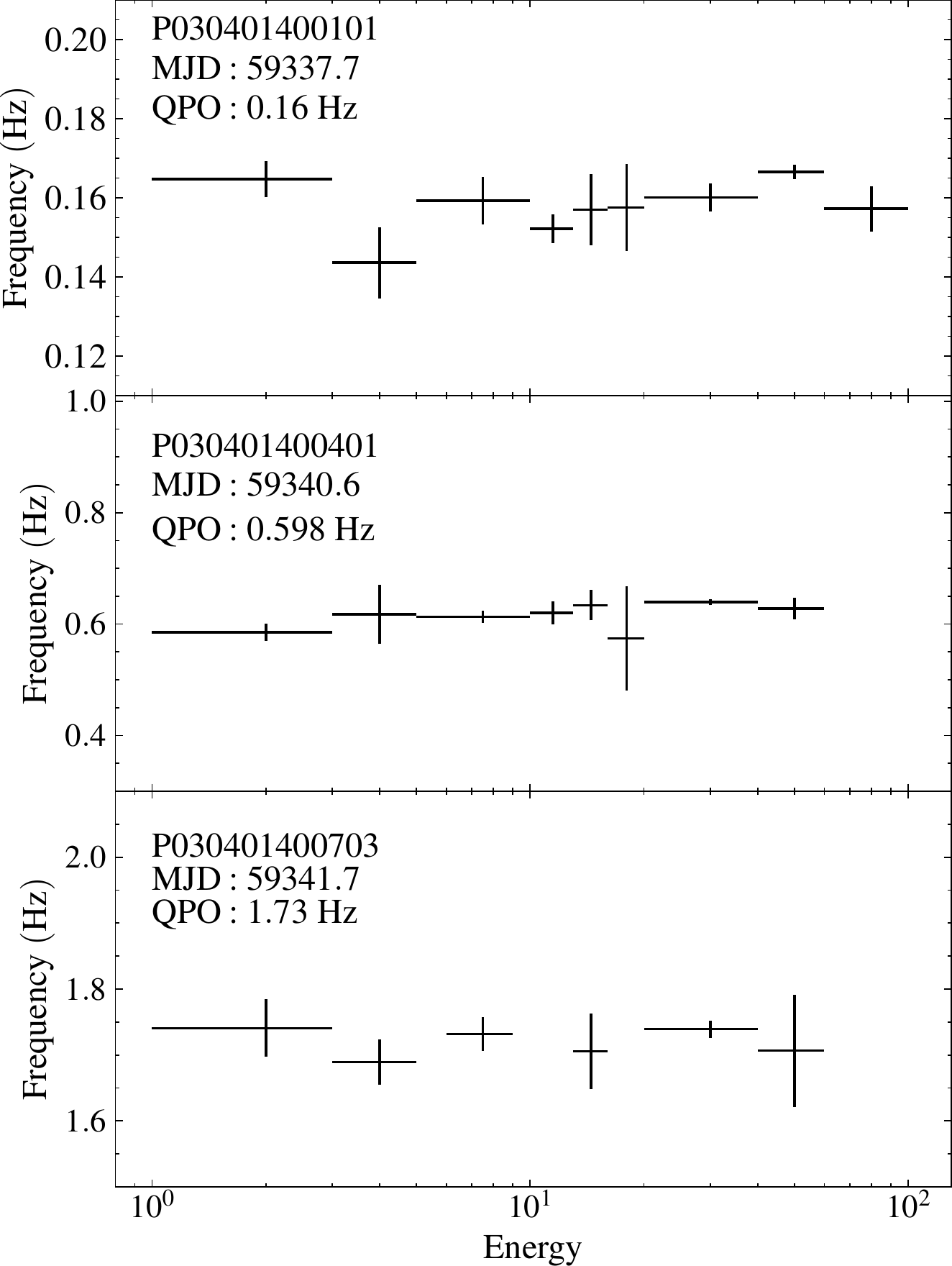}
    \caption{The frequency of Type-C QPOs as a function of photon energy.  Three representative examples were selected, 
    and the frequency remained relatively stable throughout the energy range, showing no significant 
    changes and remains fairly constant.
     }
    \label{hxmtfreqenergy}
   \end{figure}

To investigate the energy-dependent behavior of the type-C QPO properties in a quantitative manner, power spectra were extracted from several energy bands. However, we observed no apparent dependence between frequency and energy. Furthermore, when we divided the energy bands, we found that the QPO signal was weak or even absent in certain energy bands, making it difficult to perform a fit. The frequency of type-C QPOs as a function of photon energy is depicted in Figure~\ref{hxmtfreqenergy}, where the corresponding obsID and MJD are labeled in the plots. We selected three representative observations in this figure, but the results were similar for the other observations. 
The three selected observations correspond to the beginning (0.16 Hz), middle (0.598 Hz), and end (1.73 Hz) stages of the type-C QPOs.
The three selected graphs also can reflect the variation of type-C QPOs over time.  

\begin{figure}
    \includegraphics[width=\columnwidth]{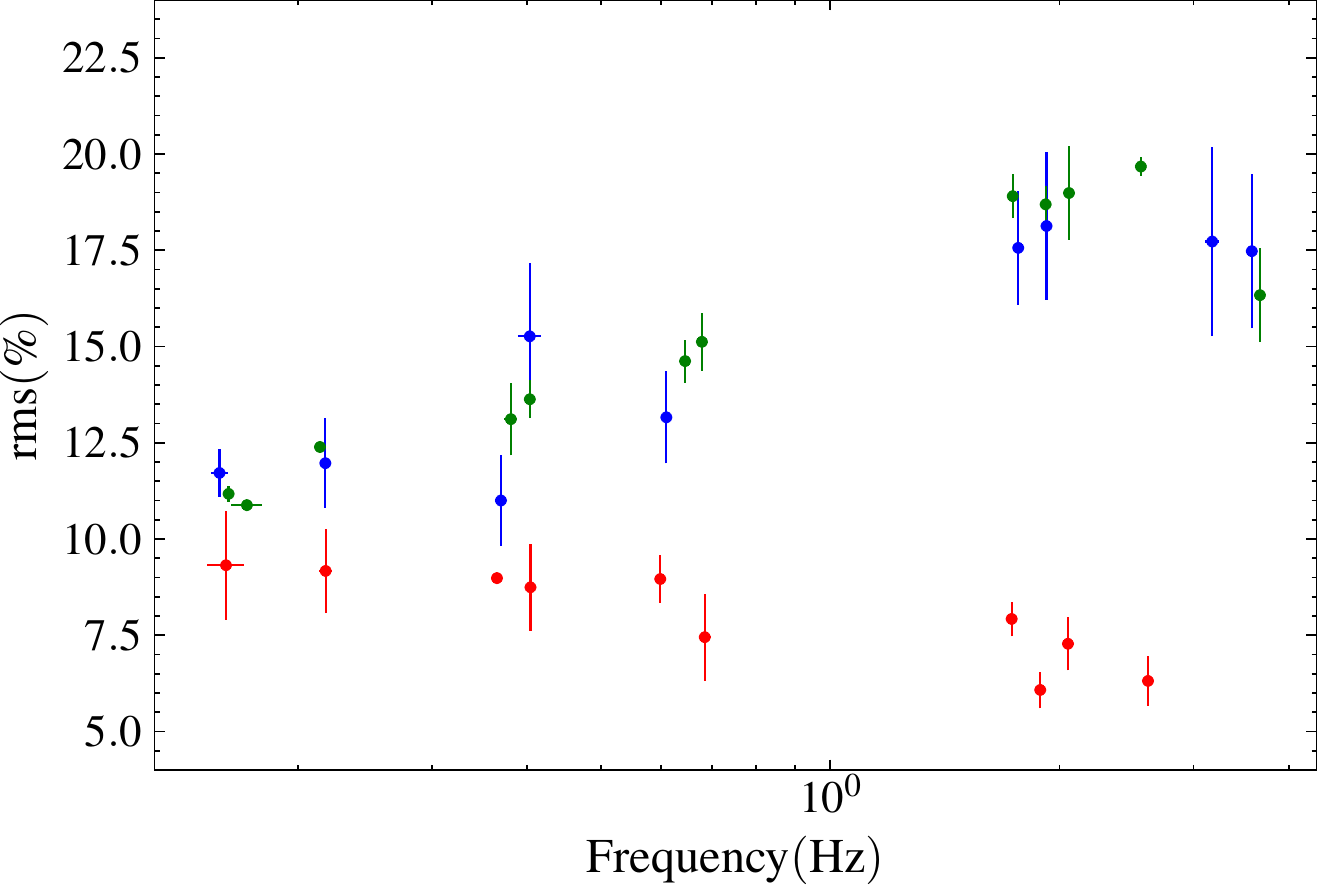}
    \caption{The QPO fractional rms as functions of type-C QPO frequencies in three detectors of Insight-HXMT: LE (red), ME (green), and HE (blue).
    The legend of the graph is consistent with that of Figure~\ref{figure3}
     }
    \label{rmsqpo}
\end{figure}

\begin{figure}
    \includegraphics[width=\columnwidth]{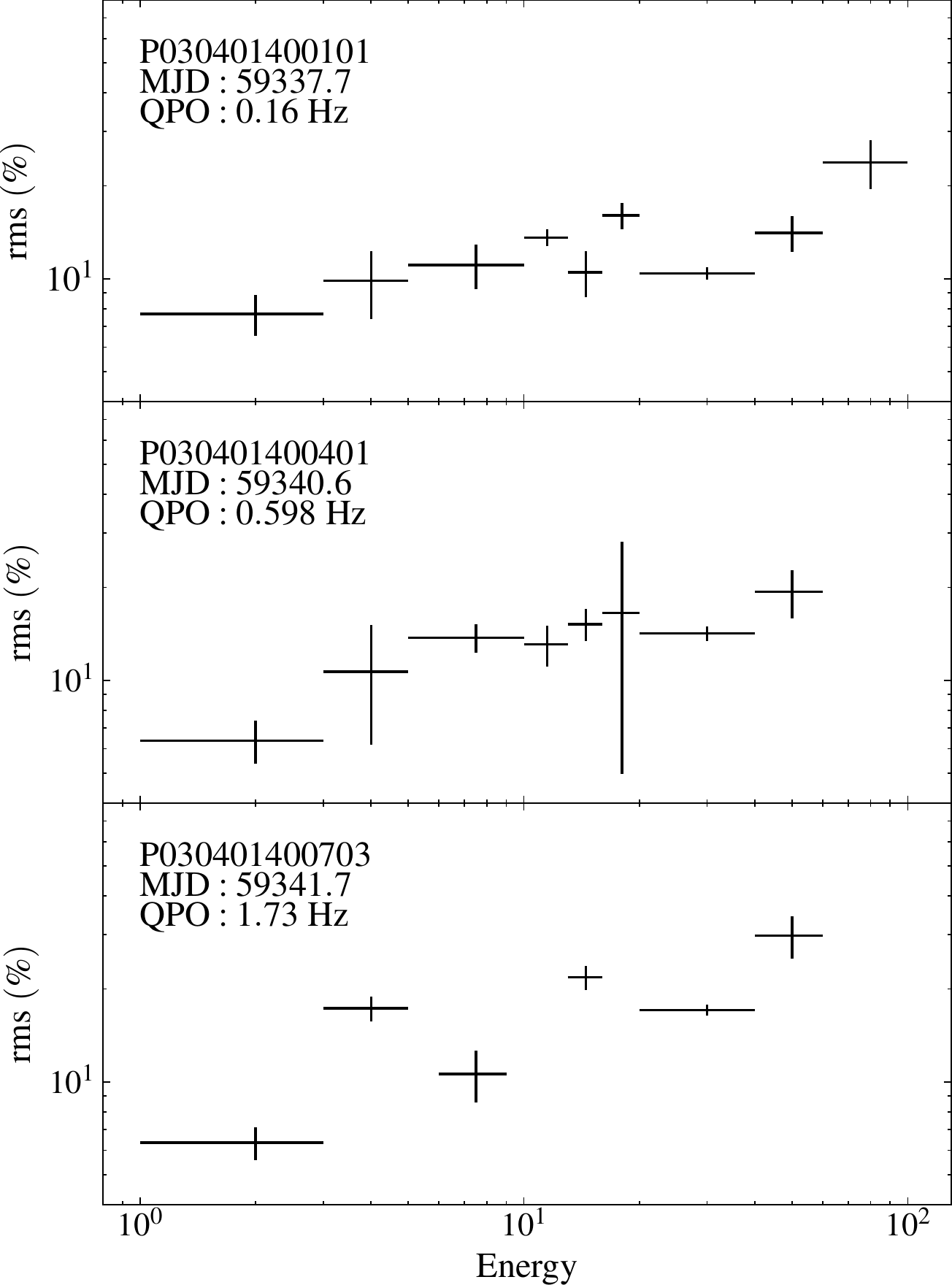}
    \caption{The fractional rms amplitude of type-C QPOs as a function of photon energy, generally the rms increases with the energies from 1 -100 keV. 
     }
    \label{hxmtrmsenergy}
\end{figure}

\begin{figure}
    \includegraphics[width=\columnwidth]{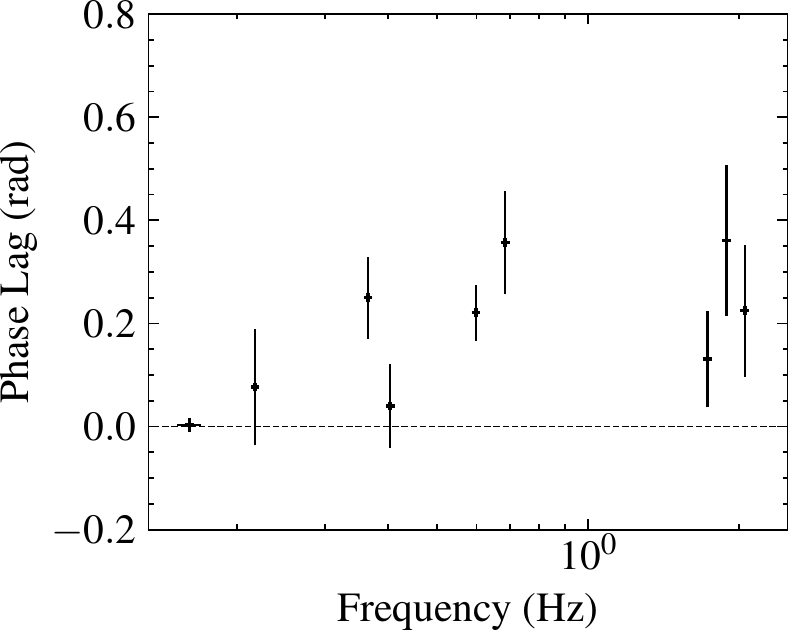}
    \caption{Phase-lags of type-C QPOs as a function of frequency. The phase lag shown in the picture displays the computed results between 3-5 keV and 1-3 keV in the observation data of \textit{insight-}HXMT. The lag is around zero from $\sim 0.1-0.4$ Hz, and above 0.4 Hz, the lag becomes positive.
     }
    \label{hxmtphafre}
\end{figure}

\begin{figure}
    \includegraphics[width=\columnwidth]{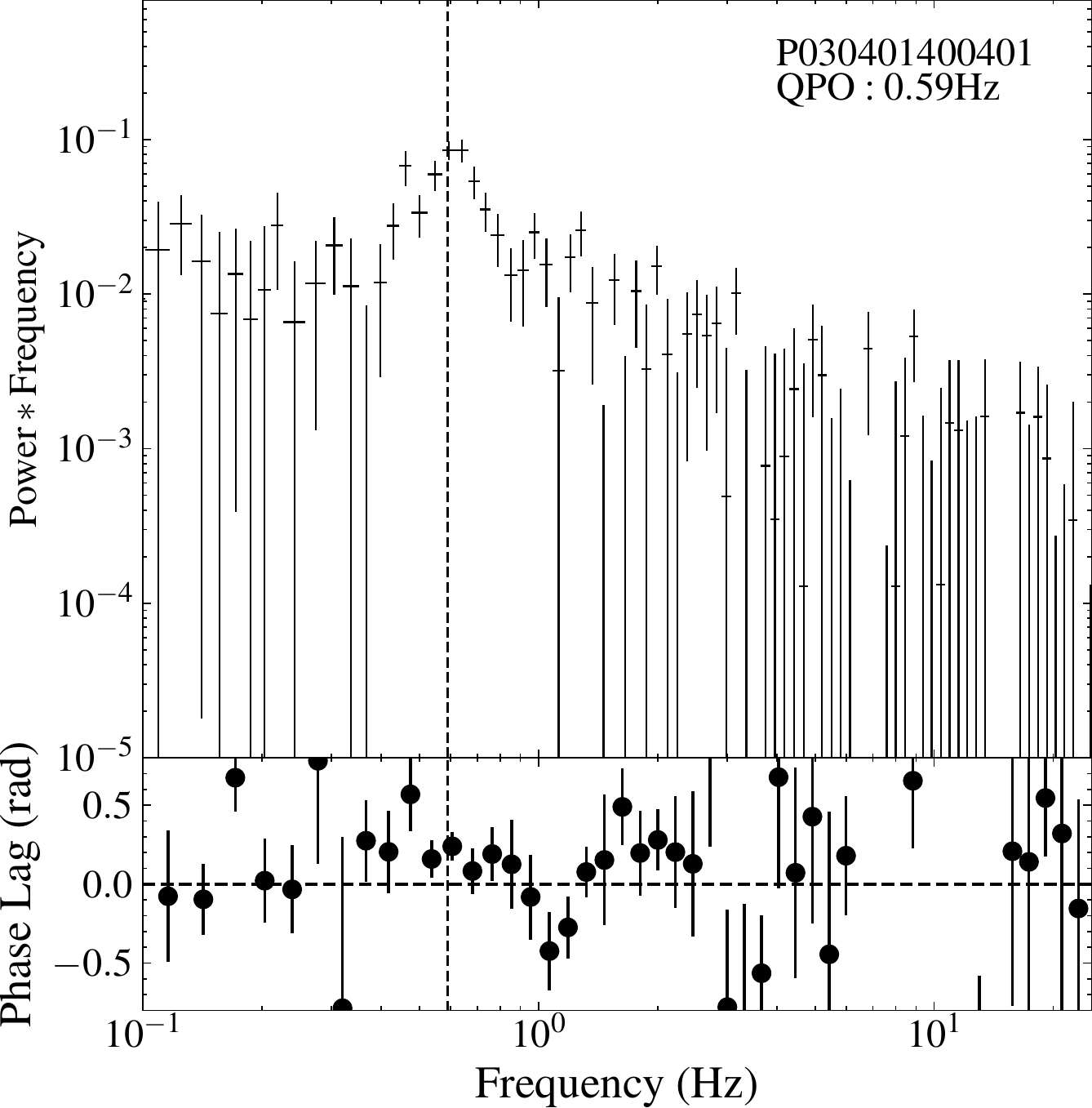}
    \caption{LE data PDS (upper) and phase-lag spectrum (lower). 
    The phase lags were calculated between the light curves corresponding to the energy 
    ranges of 1–3 keV and 3–5 keV. The frequency of the QPO is marked by the 
    dashed vertical line. For convenience, we set the rebin parameter to 0.1 in the phase-lag spectrum.
     }
    \label{hxmtphapds}
\end{figure}

\begin{figure}
    \includegraphics[width=\columnwidth]{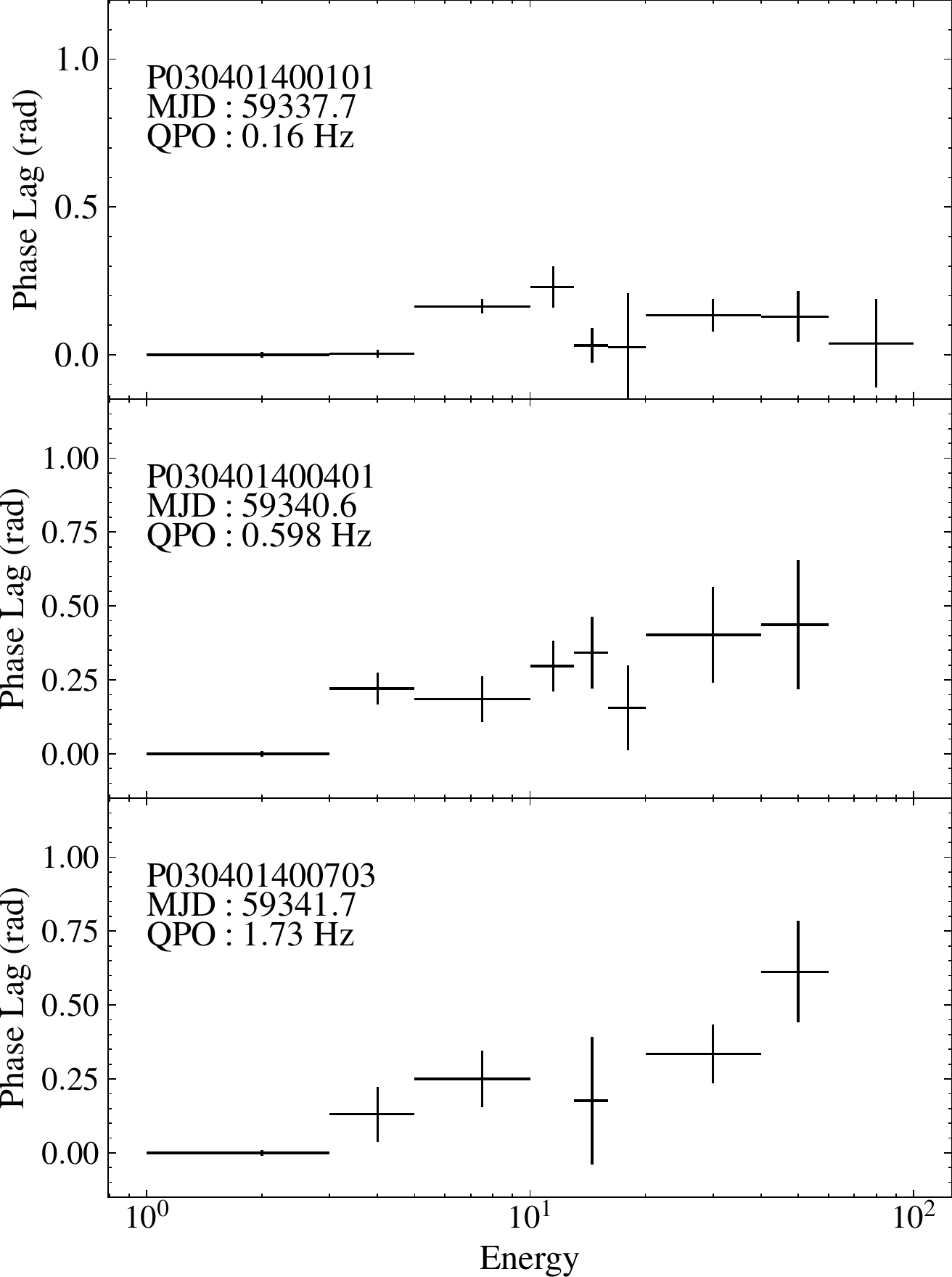}
    \caption{The representative phase-lag spectra for type-C QPOs in  MAXI J1803$-$298 from 1- 100 keV based on the Insight-HXMT observations. In the early phase of the frequency around $0.16$ Hz the lag was near zero, as the outburst evolved, the QPO frequency increased to $\sim 0.5-2$ Hz, and the lag became positive and increased gradually as the energy rises. 
     }
    \label{hxmtphaenergy}
   \end{figure}

Using data from \textit{Insight}-HXMT, we computed the fractional rms amplitude and phase-lags of the type-C QPOs as the function of both frequencies and 
energy, which are presented in Figure~\ref{rmsqpo} to \ref{hxmtphaenergy}. 
In Figure~\ref{rmsqpo}, we present the relationship between type-C QPOs' rms and frequency, and it is evident that the relationship between type-c QPOs' rms and frequency in the LE energy band is completely different from that in the ME and HE energy bands. In the ME and HE energy bands, the rms increase continuously with increasing energy, while in the LE energy band, it remains at a relatively high level at the beginning stage and gradually decreases afterward. As the frequency shows an increasing trend over time, so Figure~\ref{rmsqpo} has a similar shape to the bottom panel of Figure~\ref{figure3}. 

In Figure~\ref{hxmtrmsenergy} we have selected three representative observations consistent with Figure~\ref{hxmtfreqenergy}, and the corresponding observation numbers and QPO frequencies are indicated on the graph.  During the initial phase (top panel of Figure~\ref{hxmtrmsenergy} ),  it was observed that the rms increased with increasing energy, but the trend gradually slowed down and became rapid at high energies. In the middle panel of Figure~\ref{hxmtrmsenergy}, the rms gradually increases with energy but remains relatively stable around $E\sim 7$ keV. In the bottom panel of Figure~\ref{hxmtrmsenergy}, the spectrum is incomplete due to the lack of QPO signals in many energy bands. Nonetheless, we can still observe the general trend that the rms increase with increasing energy while maintaining relative stability. 

In Figure~\ref{hxmtphafre}, we present the phase lags as a function of frequency between the reference energy bands of 1-3 keV and 3-5 keV. Similar to the previous analysis, we excluded some observations where the signal was weak or absent after dividing the energy bands. We observed a correlation between type-C QPOs' lag and frequency, at lower frequencies, the phase lag of type-C QPOs is near 0, and then as the frequency increases, the phase lag becomes larger. However, due to the lack of observations between 0.6 and 1.7 Hz and the max frequency is around 2 Hz, we failed to notice the phase-lag sign change's transitional stage, as \cite{chand2022} calculated the time lags at 5.4 Hz and 6.3 Hz between the 3–5 and 9–12 keV energy bands are soft. 

We present in Figure~\ref{hxmtphapds} a representative 
of the phase lag as a function of frequency. The vertical dashed line denotes the frequency of the QPO, it is evident that there is a phase lag greater than zero in the vicinity of the QPO signal. In Figure~\ref{hxmtphaenergy}, we plotted the relationship between phase lag and energy for three observations.  At the beginning of the outburst, the phase lag increases slightly with increasing energy and then maintains a relatively flat shape. As the outburst progresses and the frequency gradually increases, the trend of phase lag becoming more pronounced with increasing energy becomes increasingly evident. 

\begin{table*}
    \caption{Observations of type-B QPOs in MAXI~J1803$-$298. The given error represents a 90\% confidence level. The presented fitting results are derived from the \textit{NICER}. }
    \centering
    \begin{threeparttable}
    \begin{tabularx}{0.8\textwidth}{cccccccc} 
        \hline
         Obs IDs & MJD & $\rm QPO$\tnote{f} $\rm(Hz)$ & $\rm Q$\tnote{f}&$\rm rms$\tnote{f}$\rm (\%)$& $\rm QPO$\tnote{h}$\rm (Hz)$ & $\rm Q$\tnote{h}&$\rm rms$\tnote{h}$\rm (\%)$ \\
        \hline
        4202130105 & 59352.74 & $5.64 \pm 0.20$ & $2.91 \pm 0.24$& $1.53\pm 0.14$ & $ - $&$ -$&$ -$ \\
        4202130106 & 59353.01 & $5.78\pm 0.08$ & $7.37\pm 0.25$& $0.77\pm 0.04$ & $2.59\pm 0.07$&$1.46\pm 0.15$&$1.06\pm 0.06$  \\
        4202130108 & 59355.33 & $6.45\pm 0.03$ & $7.14\pm 0.10$& $2.25\pm 0.08$ & $3.33\pm 0.07$&$2.96\pm 0.19$&$1.86\pm 0.12$ \\
        4202130110 & 59357.40 & $6.42\pm 0.04$ & $6.12\pm 0.07$&  $2.37\pm 0.07$ & $3.13\pm 0.06$&$1.59\pm 0.09$&$2.78\pm 0.13$\\
        4202130111 & 59357.98 & $7.02\pm 0.04$ & $11.80\pm 0.26$ &  $1.32\pm 0.10$ & $ - $&$ -$&$ -$ \\
        4202130112 & 59359.08 & $5.18\pm 0.07$ & $9.31\pm 0.02$ &  $1.41\pm 0.02$ & $2.77\pm 0.04$&$4.11\pm 0.06$&$1.41\pm 0.07$\\
        4675020101 & 59355.01 & $7.03\pm 0.04$ & $7.38\pm 0.14$&$1.86\pm 0.07$  & $ - $&$ -$&$ -$ \\
        4675020102 & 59356.04 & $6.75\pm 0.06$ & $7.17\pm 0.04$&$2.11\pm0.05$  & $3.24\pm 0.07$&$1.59\pm 0.02$&$2.71\pm 0.03$\\
        4675020103 & 59357.01 & $6.99\pm 0.06$ & $5.08\pm 0.10$&$2.11\pm0.09$ & $3.14\pm 0.08$&$0.91\pm 0.10$&$3.13\pm 0.12$\\
        4675020104 & 59358.18 & $7.38\pm 0.15$ & $3.81\pm 0.08$&$1.85\pm0.06$ & $2.44\pm 0.14$&$1.46\pm 0.06$&$2.44\pm 0.06$\\
        \hline
    \end{tabularx}
    \begin{tablenotes}
        \item[f] The parameters for the fundamental component are denoted by the superscript "f" 
        \item[h] The parameters for the harmonic component are denoted by the letter "h"
        
    \end{tablenotes}
    \end{threeparttable}
    \label{tab:nicerqpo}
\end{table*}

\subsection{Type-B QPOs}

\begin{figure}
    \includegraphics[width=\columnwidth]{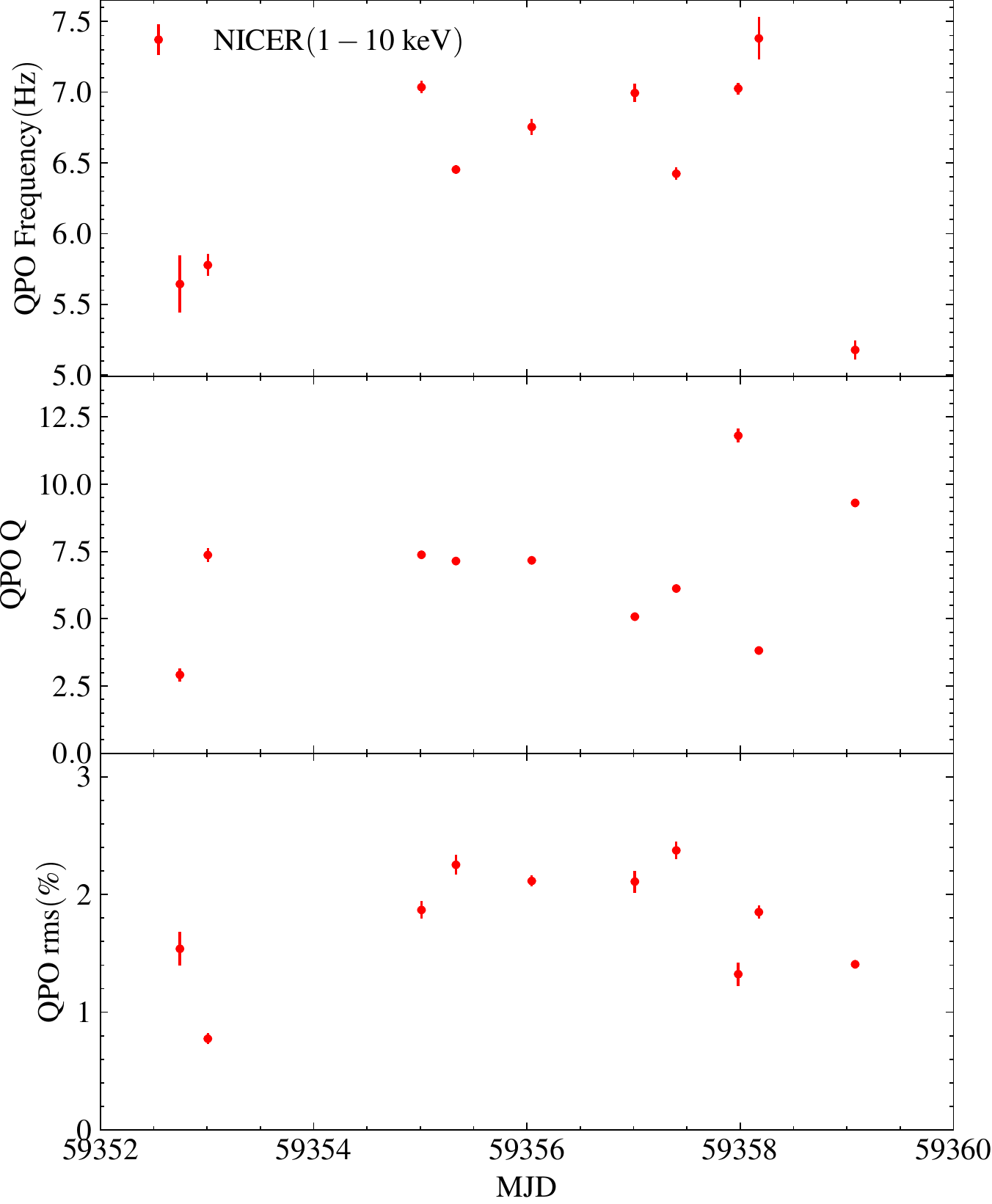}
    \caption{The evolution of the type-B QPO's frequency, Q value, and rms of the QPO with time. The data is obtained from \textit{NICER} data in the 1–10 keV energy range.
     }
    \label{nicerallqpo}
   \end{figure}

After the report by \cite{ubach2021nicer} on the detection of type-B QPOs by \textit{NICER}, we presented all the type-B QPO signals detected by \textit{NICER} in Figure~\ref{nicerallqpo}. These signals all occurred in the intermediate state. 
We roughly use the time of the first detection of type-B QPOs as the division between 
the soft and hard intermediate states. Throughout the observed period, the frequency initially increases from $\sim 5.5$ Hz to 7 Hz, 
then remains stable around 7 Hz, and finally decreases to 5.5 Hz rapidly in the last observation. We have also obtained the harmonics frequency, Q factor, and rms, and we have presented the results in Table~\ref{tab:nicerqpo}. 


\begin{figure}
    \includegraphics[width=\columnwidth]{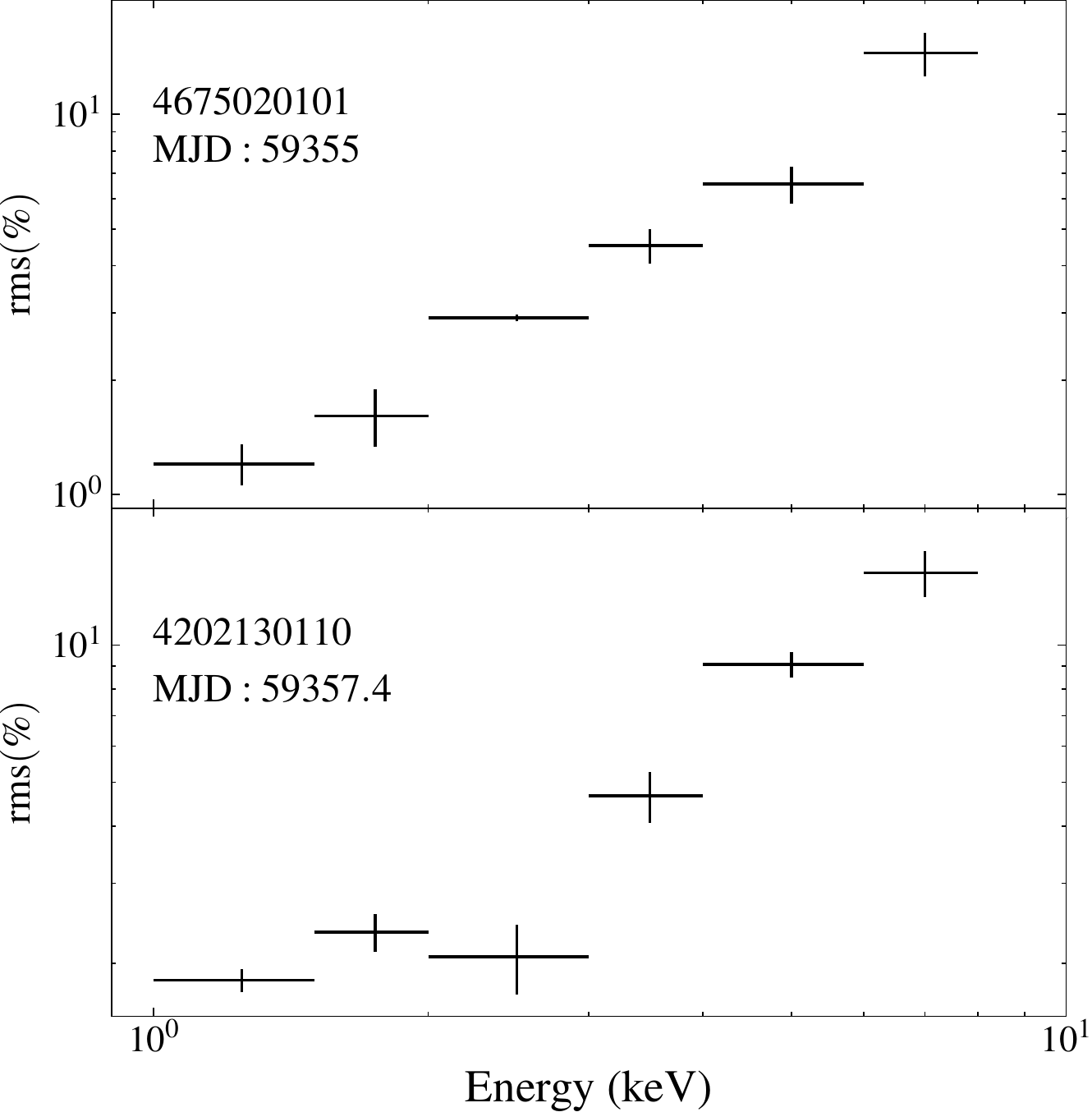}
    \caption{The representative rms spectra of the type-B QPO of MAXI J1803$-$298. The data used for this analysis were obtained from \textit{NICER}.
    }
    \label{nicerrms}
   \end{figure} 

\begin{figure}
    \includegraphics[width=\columnwidth]{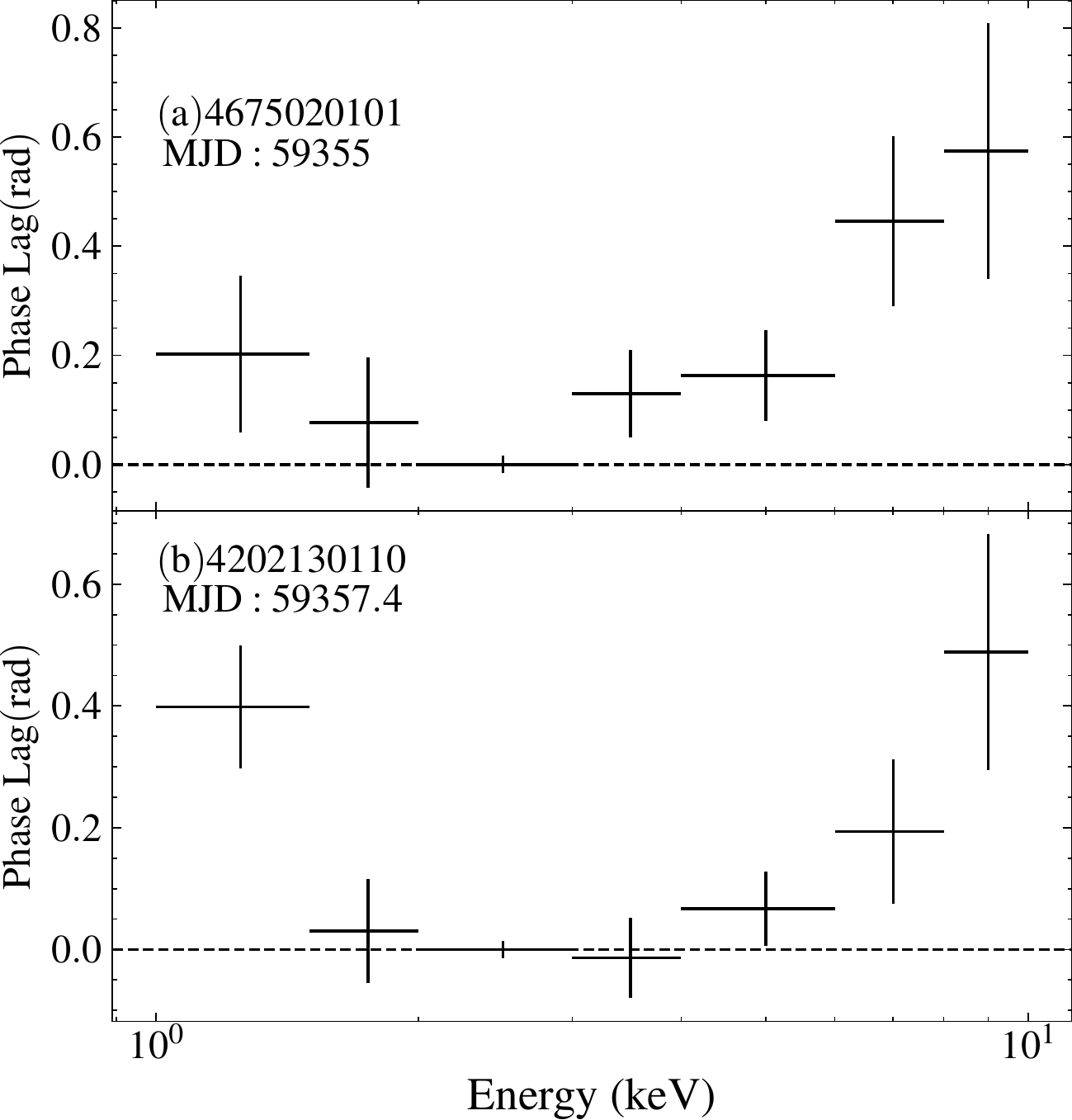}
    \caption{The representative phase-lag spectra of the type-B QPOs in MAXI J1803$-$298, which exhibit a noticeable \textit{‘U-shape’}. 
    The data used for this analysis were obtained from \textit{NICER}. 
    The horizontal dashed line represents zero phase lag.}
    \label{nicerpha}
   \end{figure} 

We present the rms and phase lags as a function of energy for the selected type-B QPOs in Figure~\ref{nicerrms} and \ref{nicerpha}, with corresponding obsIDs labeled in the figure. In the top panel of Figure~\ref{nicerrms}, the fractional rms amplitude remains at a relatively low level in the energy band of 1-2 keV, and as the energy increases, it rapidly rises to about 10\%, 
In the bottom panel of  Figure~\ref{nicerrms}, there are slight changes in the spectrum shape, and the rms value in the energy band of 2-3 keV also becomes very low. In Figure~\ref{nicerpha}, the overall phase lags shape of the spectrum shows a \textit{‘U-shaped’}, decreasing from $\sim$ 0.2 (0.4) rad at 1-1.5 keV to 0 rad at 2-3 keV (the reference band energy, but at the energy band of 3-4 keV shows a slightly negative in the bottom panel ), and increasing again above that energy to $\sim$ 0.6 (0.5) rad at 8–10 keV in the top panel (bottom panel). 

\section{Discussion}
\label{discussion}
\subsection{Outburst and Source States}
In this work, we have presented the timing results of  MAXI J1803$-$298 during its 2021 outburst using \textit{Insight-}HXMT data. The evolution during the outburst is consistent with the typical evolution of a black hole binary system 
\citep{belloni2005evolution, munoz2011black, huang2018}. During the outburst, the type-C and type-B QPOs are detected.

Based on our analysis of $Insight$-HXMT data,  we determined the state of the BH system. 
A typical state transition can be seen from Figure~\ref{figure2}, it went through the LHS, HIMS, SIMS, and HSS during the 2021 outburst. 
The rms value in the LHS that we have defined, was around 25\%, except for the initial few observations. After MJD 58349, the rms decreased to a lower level, significantly below that of the LHS, and remained stable for a long time. So we classified this period as the intermediate state. In the intermediate state, type-B QPOs appeared at about MJD~59337 indicating a transition to the SIMS. Due to the lack of the \textit{NICER} observations before this time, so we take this time as the division of HIMS and SIMS. 

The aforementioned state transitions are also consistent with the spectral fit of the MAXI/GSC and Swift/ BAT observations provided by \cite{shidatsu2022}. In their study,  the power-law photon index $\Gamma$ in the LHS is around $\sim$ 1.7,  and as the source entered the intermediate state, $\Gamma$ increased to $\sim 2.46$. At the beginning of the HSS phase, $\Gamma$ decreased to $\sim 2.1$. Due to the lack of data in the high-energy band during subsequent observations, the $\Gamma$ could not be constrained. Therefore, they obtained an acceptable fit by simultaneously fitting the soft state spectrum, which resulted in a spectral index of $\sim 2.1$. Since the energy range covered by the \textit{insight}-HXMT is wider, subsequent research can better study the transition of the source's state by fitting its energy spectrum. 

\subsection{Energy Dependence of Type-C QPO Parameters }

In the observations of \textit{Insight-}HXMT,  we observed type-C QPOs in LHS with low frequency, and without harmonic has been detected.  In Figure~\ref{hxmtfreqenergy}, we can see that the frequency of type-C QPOs does not change significantly with energy. This implies that the same QPO frequency is given at different radiation regions, which could include different parts of the jet or inner hot flow. The rms of the type-C QPOs decreased over time and QPO frequency in the LE band while increasing in the ME and HE bands as the bottom panel of Figure~\ref{figure3} and Figure~\ref{rmsqpo}. This behavior is consistent with that observed in GRS 1915+105 \citep{yan2013statistical}, EXO 1846–031 \citep{liu2021timing}, and H1743$-$322 \citep{li2013energy, shui2022trace}. 

In the previous investigation on the energy dependence of type-C QPOs, the rms amplitude of the QPO rapidly 
increased as the energy rose, but as the energy continued to rise, the growth rate gradually slowed down and tended towards flat, like MAXI J1535-571 \citep{huang2018}, XTE J1550--564 \citep{li2013energy}, and MAXI J1631-479 \citep{bu2021broadband}. Based on the Lense–Thirring (L-T) precession model, \cite{you2018x} computed the fractional rms spectrum of the QPO to study the energy-dependent variability. They found that the rms at higher energy $E>10$ keV becomes flat when the system is viewed with a large inclination angle and it is appropriate to use this explanation for the source above.   \cite{chand2022}  investigated the energy dependence of the fractional rms variation for MAXI J1803$-$298, and they identified frequencies at 5.4 and 6.3 Hz in the HIMS, as demonstrated in their Figure 5. They observed that the fractional rms variability amplitudes of both QPOs increase initially with photon energy up to $\sim$12 keV, and then either remain constant or decrease slightly. This result is consistent with those of other sources.

Although the fractional rms of this type of QPO typically increase with photon energy, it is usually observed in IMS.
But in Figure~\ref{hxmtrmsenergy}, 
we observe that the higher fractional rms amplitude at high energies and lower amplitude at low energies is consistent with the previous description. 
While the fractional rms of type-C QPOs in other systems increase up to approximately 10 keV and remain approximately constant above that energy. In our case, the energy from which the fractional rms of the type-C QPOs remain constant is not fixed, and its shape changes with time. 

After analyzing the energy spectrum of MAXI J1348-348 in \cite{zhang2020nicer}, it was found that the Comptonized component dominated the total emission during the observations with type-C QPOs. This suggests that the variability of the system is mainly due to the Comptonized component rather than the disc component, and hence the energy dependence of the fractional rms amplitude and the phase lags can be attributed to the former. In the results of our analysis on the variation of the energy dependence of the fractional rms amplitude (Figure~\ref{hxmtrmsenergy}), the shape of the spectrum varies at different frequencies, and the most likely underlying cause could be variations in the  Comptonized component. The shape of the fractional rms spectrum at different frequencies may reflect the evolving properties of the disk and Compton components around the source. Specifically, variations in the Comptonized component are likely responsible for the observed energy dependence of the type-C QPOs. Therefore, further analysis of the spectrum of this source is needed to investigate the properties and origin of the type-C QPOs. 

The physical origin of the type-C QPOs may have different models, while the L-T precession model as one of popular scenarios, occurs in a truncated disk model where a radially extended region of the hot inner flow is assumed. We have employed this model to explain the QPO features in this work, while the alternative models may also provide viable explanations for the observed phenomena. Additionally, there are some questionable issues, such as in luminous hard states ($L > 0.1 L_{\rm Edd}$), the observed X-ray spectra indicate that the accretion of the hot flow at supersonic speeds challenges conventional viscous torques, while the presence of type-C QPOs suggests that solid-body L-T precession is an unlikely underlying mechanism \citep{marcel2021can}.

Various studies have provided evidence in favor of the geometric explanation for the type-C QPOs by investigating how the timing properties change with respect to the orbital inclination angle. We have calculated the phase lag between the 1–3 and 3–5 keV energy bands using \textit{Insight-}HXMT data, and we found that the QPOs' phase lag is approximately zero and gets harder as the frequency increases. It is possible that this feature is caused by BBN. At a lower frequency of less than 1 Hz, the PDS seems to be dominated by noise components, and the lag calculation can be highly contaminated.       
\cite{chand2022} calculated the time lags as a function of energy for the QPOs at $\sim 5.4$ Hz  and $\sim 6.3$ Hz between the 3–5 and 9–12 keV energy bands are soft. 

\cite{vandeneijnden2017} discovered that the phase-lags of type-C QPOs are strongly dependent on inclination. They used 15 samples of black hole X-ray binaries and found that all samples exhibited a hard phase lag at low frequencies, while samples with low inclination remained hard lags at high frequencies, while high inclination became soft lag. The frequency and phase-lags relationships are at low frequencies with a slight hard lag in Figure~\ref{hxmtphafre}. When it comes to the high-frequency part, the time lags shows an obvious soft lag in \cite{chand2022}. Combining the behavior of QPO phase-lag at low and high frequencies, which matches the behavior of sources with high inclinations. \cite{feng2022} performed a spectral analysis of \textit{NuSTAR} and \textit{NICER} observations on the same day. 
Implementing the relativistic reflection model, they found that the source MAXI J1803–298 has a high-inclination angle, $i \thicksim 70^\circ$. The analysis results of \cite{coughenour2023reflection} using a reflection model on the \textit{NuSTAR} data also indicate that this is a source with an inclination angle of $\sim 75^\circ$. 

Several models have been proposed to explain the changing sign of the time lag of type-C QPOs. The Comptonization model proposed by \cite{nobili2000comptonization} comprises the corona consisting of two components: an inner, hot, and optically thick component, and an outer, optically thin component. This model predicts that the sign of the lags changes depending on the truncation radius of the disk and the optical depth of the corona. Specifically, at high QPO frequencies ($>$ 2 Hz), the inner disk is truncated at small radii and the inner corona is so optically thick that it up-scatters all soft photons, resulting in eventual soft lags due to down-scatterings. At lower QPO frequencies ($<$ 2 Hz), the disk is truncated at larger radii and the corona becomes optically thinner, leading to Compton up-scattering of soft photons and hard lags.

\subsection{Energy Dependence of Type-B QPO Parameters}

Type-B QPOs were observed from \textit{NICER} observations in the IMS. Our fitting results showed that its frequency was 5$\sim$ 7.5 Hz, consistent with its typical frequency in BHXBs. Type-B QPOs are typically detected during a transition of a source to the SIMS, like MAXI J1535$-$571 \citep{huang2018} and GX 339-4 \citep{belloni2005evolution}. Type-B QPOs and the X-ray flux peak are believed to be associated with jet ejections \citep{fender2009jets}. Our  
data in Figure~\ref{figure1} indicates that the occurrence of QPOs corresponds to the X-ray flux peak. However, the relationship with jets needs to be further verified with multi-wavelength data. \cite{wood2023time} preliminarily confirmed the existence of a jet during the peak of a radio flare at this source using a time-dependent model, but the exact time needs to be further confirmed. For MAXI J1803$-$298, observations in the radio may confirm the relationship between type-B QPOs and jets in the future. 

For the type-B QPO, we studied the relationship between the rms and energy.  As we can see, the rms of the type-B QPO gradually increase as the energy rises. This behavior is similar to that observed in many other sources that have been studied, like MAXI J1348-630 \citep{belloni2020time,garcia2021two,bellavita2022vkompth} and GRS 1915+105 \citep{garcia2022evolving}. 

The results of the relationship between the phase lags and 
energy of type-B QPOs obtained from the observation data of \textit{NICER} are shown in Figure~\ref{nicerpha}. 
The phase lags spectrum is '\textit{U-shaped}', similar shape is also found in MAXI J1348-630 \citep{belloni2020time,garcia2021two}, 
GX 339-4 \citep{peirano2023dual}, MAXI J1820+070 \citep{ma2023detailed}. Furthermore, the shape of the phase lag spectrum also varies with time, and in other observations that we did not present, there was not a very typical shape like this, all of which reflect the evolutionary characteristics of the source.

The mechanism to explain the hard lag is the Comptonization of soft photons in the hot corona. \cite{belloni2020time} proposed that the soft lags at energies below 2 keV are caused by Compton down-scattering in the corona of the photons emitted by the disc 
using a flat seed photon spectrum emitting exclusively between 2 and 3 keV. But if the reference photon has zero lag, 
then both high-energy and low-energy photons will exhibit positive hysteresis. 

Based on the time-dependent Comptonization model of \cite{karpouzas2020comptonizing} and  \cite{garcia2021two} introduced two independent Comptonization regions (lacks a feedback interaction between the two Comptonization regions), then they got a good fit when fitting the phase-lag spectra of the QPO. They demonstrated that the 'U-shaped' phase lag spectrum can be described by accounting for both the soft lags at low energies and the hard lags at high energies, which are explained by photons from the disc being inverse-Compton scattered in the corona.
\cite{bellavita2022vkompth} modeled the accretion disc as a 
multi-temperature blackbody source generating soft photons that are subsequently Compton up-scattered in a spherical corona, with feedback from Comptonized photons that return to the disc. Further analysis is required to understand the type-B QPO's origin and geometry of the Comptonizing region. 

In the previous work on the fitting of similar rms and phase lag spectra, single-corona models were first used, but the obtained results could not satisfactorily describe the spectral shapes, particularly in energy ranges before about 2 keV and after 8 keV. Therefore, they all resorted to the dual-corona model to fit the spectra \citep{garcia2021two,ma2023detailed}. Eventually, they obtained highly consistent fitting results and reasonable physical parameters. The shapes displayed in the rms and phase lag spectra in the two observations we presented in our paper are very similar to those from previous fitting studies of several sources. We infer that a single corona model cannot fit the data well, and a dual corona model should be able to provide better-fitting results and physical interpretations. 

\section{Conclusion}
\label{con}
We have presented a timing analysis of the new  black hole candidate MAXI J1803$-$298 
using \textit{Insight-}HXMT and \textit{NICER} observations, and reported the detection of both Type-C and Type-B QPOs during the 2021 outburst. The main results of the study are summarized as follows:
\begin{itemize}
    \item The source evolved from LHS to HIMS, then to SIMS, and finally to HSS. 
    \item Due to the wide X-ray bands based on the \textit{Insight-}HXMT observations, we were able to analyze the energy dependence of the type-C QPO fractional rms and frequency of MAXI J1803$-$298 up to 100 keV for the first time. There is no significant variation in the frequency with energy. The energy dependence rms is consistent with other sources. The relationship between QPO rms and energy is consistent with the geometric origin of type-C QPOs.
    \item Considering the phase lag between the source's type-C QPOs at low and high frequencies, as well as their relationship with frequency and energy, it implies that MAXI J1803$-$298 is a high-inclination source. 
    \item By comparing the rms and phase lag spectra of the type-B QPOs with previous studies, it is suggested that a dual-corona model would offer a better physical interpretation.

\end{itemize}
\section*{Acknowledgements}
We express our gratitude to the referee for their insightful comments and valuable suggestions, which have greatly enhanced the quality of the paper.
This work is supported by the National Key Research and Development Program of China (Grants No. 2021YFA0718503), the NSFC
(12133007, U1838103). This work has made use of data from the \textit{Insight-}HXMT mission, 
a project funded by the China National Space Administration (CNSA) and the Chinese Academy of Sciences (CAS).
\section*{Data Availability} 
Data that were used in this paper are from the Institute of High Energy
Physics Chinese Academy of Sciences(IHEP-CAS) and are publicly available for download from the 
\textit{Insight-}HXMT website \textbf{\url{http://hxmtweb.ihep.ac.cn/}}. The data of \textit{NICER} used in this article can be obtained from this website \url{https://heasarc.gsfc.nasa.gov/cgi-bin/W3Browse/w3browse.pl}. 



\bibliographystyle{mnras}
\bibliography{1}


\bsp	
\label{lastpage}
\end{document}